\documentclass[12pt]{article} 
\usepackage{graphicx,amsmath,amsfonts}

\def\fun#1#2{\lower3.6pt\vbox{\baselineskip0pt\lineskip.9pt
  \ialign{$\mathsurround=0pt#1\hfil##\hfil$\crcr#2\crcr\sim\crcr}}}

\newcommand{\be}{\begin{equation}}
\newcommand{\ee}{\end{equation}}
\newcommand{\ba}{\begin{eqnarray}}
\newcommand{\ea}{\end{eqnarray}}
\newcommand{\bg}{\begin{gather}}
\newcommand{\foma}{\end{gather}}

\title{ Excitation of Physical Vacuum}

\author{
M.K. Volkov$^a$, E.S. Kokoulina$^{a,b}$ and E.A. Kuraev$^a$ 
\vspace{4mm}
\\
\small\sl $^a$ Joint Institute for Nuclear Research, Dubna, 141980, Russia\\
\small\sl $b$ Gomel State Technical University, Belarus
}

\date{}
\begin{document}
\maketitle

\begin{abstract}
Introducing such a notion as a "excitation of physical vacuum"  we do an attempt to
explain some strange experimental facts such as large value of $\sigma$-term
measured in pion-nucleon low energy scattering, $\Delta T=1/2$ rule in kaon two pion decay modes,
the ratio of strange to nonstrange yield in low energy proton-antiproton annihilation,
excess of soft photons in hadron's collisions. As a test of our approach we suggest to 
measure the multiparticle production processes and decays of heavy virtual objects.
\end{abstract}

\section{Motivation}

Motivation of this note is the need to explain some strange experimental facts
in particle physics.

One of them is too large value of strangeness content $y=<P|\bar{s}s|P>/<P|\bar{u}u+\bar{d}d|P>$
in the nucleon $|P>$ state. This quantity can be expressed with the experimentally measured 
value of so called $\sigma$- term:
\ba
\sigma=\frac{\sigma_0}{1-y},
\ea
with $\sigma=80-200 MeV$-the measure of chiral symmetry violation 
and can be expressed in terms of pion-nucleon scattering lengths,
$\sigma_0\approx 30 MeV$-can be calculated in frames of quark-parton model
(\cite{ChengLee}, Chapter 5). It is difficult to explain too large value of $y>0,5$. 

Another one is the experimental evidence about "too large strangeness" in proton 
which follows from the experiments
with annihilation of proton and antiproton near threshold into mesons \cite {Sapoz}.
It turns out that in the case when initial particles are in state $^3P_0$
the ratio of kaon production to pion production is of order of unity:
\ba
\frac{Br(\bar{p}p\to KK)}{Br(\bar{p}p\to \pi\pi)}\approx 1
\ea  
contrary to the state $^3S_1$ where this ratio is small.

We mention here the problem of understanding of an excess of soft photons
created at high energy hadrons collisions \cite{SF}. There are some 
uncertainties in model description of high multiplicity processes \cite{HM}.
As well we mention here the long-standing problem of explanation of $\Delta T=\frac{1}{2}$
rule in weak kaon decays.

\section{Physical vacuum excitation}

In this note we do an attempt to understand these phenomena in frames of 
model of the Physical Vacuum (PV)
excitement \cite{VKK}. We regard PV as a Dirac cellar which can be excited when accepting some
amount of energy released during the collision of initial particles (or decay of 
heavy initial particle). This excited state is a state containing any number of gluons 
and light (current) quark-antiquark pairs with quantum number of vacuum $0^{++}$.
We suppose the equal probabilities of presence of quark-antiquark germs of any flavor.
PV excited state is similar to the state of a liquid with the temperature close to 
the boiling point. When the germs accept sufficient amount of energy the current 
quarks turn out to the constituent ones and at the hadronization stage reveals 
itself as a mesons or nucleons. For the case proton-antiproton annihilation the
branching ration of kaon and pion production is expected to be approximately equal.

For collisions with total energy large enough to produce the charm germs or the 
ones of beauty the rates of production of corresponding hadrons  (when taking into 
account the difference of phase volumes) becomes of the same order. From this point 
of view the experimental result (2) can be accepted. As for (1) we expect the PV
contribution to $y=y_{PV}+y_I$ dominate $y_{PV}=0,5$ for the case of proton-antiproton 
annihilation whereas the contribution of the intrinsic one $y_I=0,15$ is more realistic
one. In paper \cite{sigma} the value of $\sigma\approx 80 MeV$ was obtained in frames 
of Nambu-Iona-Lazinio model with $\sigma$-pole intermediate state.

The mechanism of the meson creation from the region of the excited vacuum state of the size $L$
remind the process of vapor bubbles creation in a hot water \cite{Andreev}. When the temperature
do not exceed the boiling point $T<T_c$ the bubbles which are always exist in a liquid 
due to fluctuations do not increase. The superficial tension dominate. The situation changes 
in the boiling point $T=T_c$: the new phase becomes more convenient energetically
and the number of the bubble of size $R$ creation become dominate:
\ba
n(R)\sim exp(-\Delta W(R)/T)\sim exp(-a(T-T_c)R^3),
\ea
with $\Delta W=\Delta E-T\Delta S+P\Delta V$-free energy, $\Delta W=\frac{4\pi}{3 v}R^3
(\mu_1(T)-\mu_2(T))+4\pi\sigma R^2$, $\mu_{1,2}$ are the chemical potentials of liquid 
and gas phases, $v$, $\sigma$-the volume per one molecule and the superficial tension.
It is important to note that the probability of creation of a large bubble (hadron in a
gluon liquid) have a resonance form:
\ba
W\sim \int r^2exp(-a(T-T_c)r^3)dr\sim \frac{1}{|T-T_c|}=\frac{1}{|M-E|}.
\ea
This expression remind the Breit-Wigner form and, presumably, can be confirmed taking 
into account the dynamics of this process.
We can conclude that any process with possible PV intermediate state will have some
enhancement, which can be associated with intermediate state with quantum numbers
$I^G(J^{PC})=0^+(0^{++})$, so called $\sigma$-meson.
The matrix element of process with production some state $X$ with vacuum quantum numbers
will have a form
\ba
M(ab\to c X)=\frac{1}{s_1-M_\sigma^2+i\Gamma_\sigma M_\sigma}M(ab\to c+ PV)M(PV \to X)+... ,
\ea
with dots denoting the contributions which do not contains the $\sigma$-pole 
contribution; $s_1$-is the invariant mass square of the set of particles $X$.

We remind that in this way the remarkable enhancement of decay amplitudes with 
$\Delta T=1/2$ in two pion modes of kaon decays can be understood \cite{Volkov}.

Consider now the phenomenon of soft photon emission at hadrons collisions.
According to our model the hot quark-gluon system appears as a result of energy 
accepted by PV. Germs consisting from the current (light) quark-antiquark pairs 
of different flavors turns to pairs of constituent (heavy) quarks and gluons which 
can be considered as almost equilibrium system. Almost real hadrons can be 
emitted from the boundary of hot vacuum surface as a real hadrons. 
This scenario is similar to Big Bump one resulting our Universe creation. In 
the last case the equilibrium between electrons, protons and photons breaks at 
temperature $T=3000^0K$ when the electron-proton recombination and the creation 
of a neutral atoms take place. At this stage photons cease to interact with the 
matter and start to expand as a relic ones. At the contemporary moment we observe 
the spectrum of relic photons (black body spectrum)\cite{Heer}:
\ba
\frac{d\rho}{d\nu}=\frac{8\pi}{c^3}\frac{\nu^2}{e^{\frac{h\nu}{T}}-1}
\ea
with $d\rho/{d\nu}$ - the spectral spatial density of photons, $c, h\nu$ - are the 
light velocity and the energy of photon. Maximum of spectral distribution is located 
at $h\nu_0=2.8k_B T$. The temperature and the density of contemporary relic photons
are $T_0=3^0K$ and $480 cm^{-3}$, correspondingly.

Massless gluons interacting with quarks  similarly to photons have a zero chemical 
potential and therefore obey the black body emission spectrum. Gluon density at
the deconfinement temperature $T_c\approx 200 MeV$ can be estimated as
\ba
\rho_{gl}=2.4(\frac{T_c}{T_0})^3*10^{-37}(fm)^{-3}\approx 0.3(fm)^{-3}.
\ea 
So the number of gluons in the hot vacuum region of size $L=20 fm$ will be  
of order of a few hundreds. What is the fate of them at the hadronization stage?
They can not be emitted as a free particles due to their open color. Besides
they can not create the colorless glueball state as well as they are mostly soft.
The gluon excess are accepted by the quarks which are now heavy. These hadrons
are almost on the mass shell. The energy excess is emitted by means of soft photons.

Dynamics of turning the soft gluons to soft photons is presumably as complicate as
the confinement phenomenon one. We will not touch it here.

Averaging on the temperature of quark-gluon system from the beginning stage
$T>>T_c$ up to $T=T_c$ at the final hadronization stage one can estimate the
soft photon emission spectrum:
\ba
\frac{d W}{d \nu}\sim \int\limits_0^{T_c^{-1}}d\beta f(\beta) exp(-\beta h\nu)\sim 
\frac{A}{\nu}f(0).
\ea
The average energy of gluons can be estimated from the total energy deficits
carrying by soft photons $\Delta E \sim 10^{-2}E$ and the number of them estimated above.
For the case of proton antiproton annihilation we estimate $h\nu\sim 1-2 MeV$.
The spectrum behavior is the same as QED soft photon emission one, but the quantity
A can be an order large the QED one \cite{Boyan} 
$A_{QED}=\frac{\alpha}{\pi}\sim 2.5\ast 10^{-3}$.

\section{High multiplicity process}

Let now consider the manifestation of PV mechanism in high multiplicity
processes. For definiteness let first consider the decay of particle
with rather high mass. Such as $\rho\to 4 \pi;J/\Psi\to 2\pi \mu\bar{\mu}$.
Matrix element will have an additional term which corresponds to
2 pions created by PV and the off mass shell meson decaying 
by 2 particles channel. 

The quantum numbers of the set of pions created from PV have vacuum 
quantum numbers $0^{++}$ and, besides, zero total 3-momentum in the rest 
frame of decaying particle. The last feature is specific one for PV
excitation mechanism.

It is easy to see that the corresponding contribution to the total width
do not contain an interference term of traditional matrix element of $n$
particles decay and $n-2$ decay matrix element with 2 pions created by 
PV mechanism. 
Really -the kinematics of creation are quite different.

So in general the width will have a form of  some finite sum of contributions
with 2 pions,4 pions and so on created through PV excitation channel.

To obtain some definite predictions we must work in frames of model
(we want it to be called "a corrupt model"(CM)).
We suggest the effective Lagrangian which takes into account PV excitation
can be build starting from the usual one $L_0$ as \cite{W}
\ba
L_{eff}=L_0[1+c_2\frac{\vec{\pi}^2}{f_\pi^2}+c_4(\frac{\vec{\pi}^2}{f_\pi^2})^2+...],
\ea
with $f_\pi$-pion decay constant.
So the matrix element of decay process of heavy object $A$ to set of particles $a$,
accompanied by emission of set of $2n$ pions $A(P_A)\to a +2n\pi$, some of them 
created through the PV excitation mechanism will have a form (see Fig. 1):
\ba
<A,PV|L_{eff}|a,2n\pi>=M(A\to a+2n\pi) \\ \nonumber
=\sum\nolimits_{k=0}^{n}M_k(A\to a_k+2k\pi)
M_0(a_k\to a+2(n-k)\pi) \\ \nonumber
+ \sum\nolimits_{k=0}^{n}M_0(A\to a+\sigma_k+2(n-k)\pi)
M_k(a_k\to 2k\pi). 
\ea
where $|a_k>,|\sigma_k>$ -are the  intermediate state with quantum numbers of $A$
and vacuum respectively.

To take into account energy distribution between sets of final-state particles,
we will write the phase volume in form:
\ba
d\Phi_{a,2n\pi}=\delta^4(P_A-\sum p_a -\sum\nolimits_{i=1}^{2n}q_i)\Pi\frac{d^3p_a}{2\epsilon_a}
\prod\nolimits_{i=1}^{2n}\frac{d^3q_i}{2\omega_i} \\ \nonumber
=d^4q\delta^4(P_A-q-\sum p_a -\sum\nolimits_{i=1}^{2(n-k)}q_i)\prod\nolimits_{j-1}^{2(n-k)}\frac{d^3q_i}{2\omega_i}
\prod\frac{d^3p_a}{2\varepsilon_a}\ast \delta^4(q-\sum\nolimits_{j=1}^{2k}q_i)\prod\nolimits_{j=1}^{2k}\frac{d^3q_j}{2\omega_j}.
\ea
Keeping in mind the kinematical fact of absence of interference terms of different contributions to
the matrix element the width can be put in form (we imply the rest frame of decaying particle):
\ba
\Gamma(A\to a,2n\pi)(M_A)=\Gamma_0(A\to a,2n\pi)(M_A) \\ \nonumber
+\sum\nolimits_{k=1}^{n}\lambda_k\int\limits_{a(k)}^{b(k)}q_0^3P_k(q_0)dq_0[
\frac{\Gamma_0(A\to a+\sigma_k+2(n-k)\pi)(M_A)}{(q_0^2-M_{\sigma_k}^2)^2+M_{\sigma_k}^2\Gamma_{\sigma_k}^2} \\ \nonumber
+\frac{(1-\frac{q_0}{M_A})\Gamma_0(A\to a+2(n-k)\pi)(M_A-q_0)}{((M_a-q_0)^2-M_{a_k}^2)^2+M_{a_k}^2\Gamma_{a_k}^2}].
\ea
with $\Gamma_0(M)$-is the width calculated in frames of standard field theory.
Here we denote $a(k)=2km_\pi,b(k)=M_A-m_a-2(n-k)m_\pi$, and the phase volume of
$2k$ particles in the nonrelativistic case \cite{Kopylov}:
\ba
\int\delta(q_0-\sum \omega_i)\delta^3(\sum\nolimits_{i=1}^{2k}\vec{q}_i)\Pi\frac{d^3q_i}{2\omega_i}=
c_k((q_0-2km_\pi)/f_\pi)^{3k-\frac{5}{2}}=c_kP_k(q_0).
\ea
We use the realistic assumption $M(A\to a_k+2(n-k)\pi)=M(\sigma_k\to 2k\pi)=M_k$.
To realize the requirement of zero total 
3-momentum of set $2k$ pions created through PV excitation mechanism, we use the
(model-dependent) assumption:
\ba
|M_{2k}|^2=q_0^3\lambda_k c_k^{-1}\delta^3(\vec{q}).
\ea
The quantities $M_k,\Gamma_k,\lambda_k$ can be considered as a free parameters.

For illustrations let consider 4-particles channels decays of $\rho$ and $\Psi'$:
\ba
\Gamma(\rho\to 4\pi)=\Gamma_0+\Delta\Gamma_\rho,
\ea
with $\Gamma_0$ can be found in \cite{ESK}(and the references therein) and (see Fig. 2)
\ba
\Delta\Gamma_\rho=\lambda_1\int\limits_{2m_\pi}^{M_\rho-2m_\pi}
q_0^3dq_0\beta(q_0)[\frac{(1-\frac{q_0}{M_\rho})\Gamma_{\rho\to 2\pi}(M_\rho-q_0)} 
{((M_\rho-q_0)^2-M_\rho^2)^2+M_\rho^2\Gamma_\rho^2} 
+\frac{\Gamma_{\rho\to 2\pi+\sigma}(M_\rho)} 
{(q_0^2-M_\sigma^2)^2+M_\sigma^2\Gamma_\sigma^2}] \\ \nonumber
\beta(q_0)=\sqrt{1-\frac{4m_\pi^2}{q_0^2}},\Gamma_{\rho\to 2\pi}(M)=\frac{Mg_{\rho\pi\pi}^2}{48\pi}(1-\frac{4m_\pi^2}{M^2})^{3/2}.
\ea
For $\Psi'\to 2\pi\mu\bar{\mu}$ decay we have $\Gamma=\Gamma_0+\Delta\Gamma_\Psi$ 
with (see Fig. 3)
\ba
\Delta\Gamma_\Psi=\lambda_1\int\limits_{2m_\pi}^{M_{\Psi'}-2m_\mu}q_0^3dq_0\beta(q_0)[
\frac{(1-\frac{q_0}{M_{\Psi'}})\Gamma_{J/\Psi\to \mu_+\mu_-}(M_{\Psi'}-q_0)}
{((M_{\Psi'}-q_0)^2-M_{J/\Psi}^2)^2+M_{J/\Psi}^2\Gamma_{J/\Psi}^2} 
+\frac{\Gamma_{\Psi'\to \sigma+\mu_+\mu_-}(M_{\Psi'})}
{(q_0^2-M_\sigma^2)^2+M_\sigma^2\Gamma_\sigma^2}]. \\ \nonumber
\ea
We remind when describing the $e^+e^-$ annihilation experiments, besides 
the possibility of vacuum excitation considered above
for energies exceeding the $\rho$ meson mass, the additional contributions
connected with higher meson resonance excitation must be taken into 
account \cite{Ivanch}.  

Model suggested will give the same result as in the traditional 
field theory for such a processes as $e^+e^- \to \pi^+\pi^-\pi^0;
K^+K^-\pi^0;a\bar{a}\gamma $. 

Really in our model in the center of mass frame of the initial 
electron and positron the sum of 3-momentum of two particles created by vacuum 
excitation  is zero as well as the 3-momentum of the third particle.In this 
kinematics the PV-matrix element equals zero.

The events with possible groups of the final particles with quantum numbers 
of vacuum and with the zero 3-momentum in the rest frame of decaying particle
can be searched at colliders with creation of large number particles such as
proton-proton or nucleus-nucleus collisions, DIS-experiments.

\section{Discussion}

The idea of PV excitation through energy transfer to Dirac cellar remind the idea 
of using tadpoles in sixteenth to explain $SU(3)$-symmetry violation \cite{CG},
so it is not new one.

The effective Lagrangian used above to describe multipion production was first 
suggested by S. Wienberg \cite{W} in frames of current algebra and further was 
derived in models with nonlinear realization of symmetry \cite{V}.

PV excitation state $|\sigma_k>$ with quantum numbers of vacuum
$J^PC=0^{++}$ can exist as a particle with definite mass and 
width and, besides, as a some object which can not be interpreted as a
definite particle. Situation remind the pomeron-state with quantum numbers
$ J^{PC}=1^{++}$ used to describe high energy hadron hadron scattering,which
can not be associated with the definite particle.

The dominant role of gluons in creation of jets at hadrons collisions was 
underlined in papers of Carruthers and E. V. Shuryak \cite{Shur}.

In paper of one of us (MKV) a reason why the ration of $K\to 2\pi$ decays
matrix elements $M(K_s\to \pi^+\pi^-)/M(K^+\to\pi^+\pi^0)\sim 50$ was given:
really in the neutral kaon decay the intermediate state with vacuum quantum 
numbers with mass value close to kaon mass was shown to give the dominant
contribution \cite{Volkov}.

Formulae sited above have a rather qualitative meaning.

\section{Acknowledgements} 

We are grateful to RFFI grant 03-02-17077 for supporting this work.
As well we are grateful to SCAR JINR and personally to Valentin Samoilov 
for the support and stimulation.

\newpage
\begin{figure} 
\begin{center}
\includegraphics[scale=1.0]{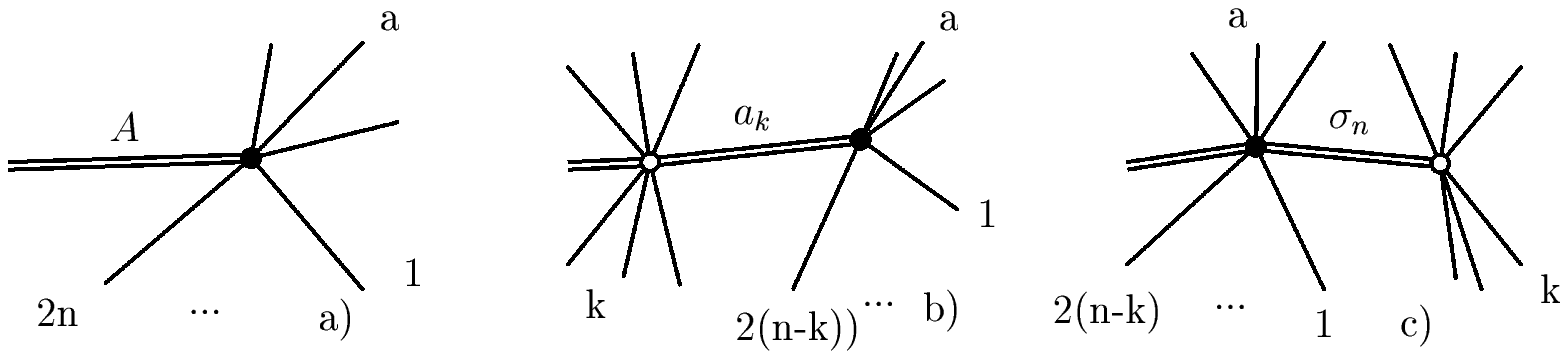}
\end{center}
\caption{ Contribution to decay $A\to a+2n\pi$
matrix element: 
a) without PV excitation,
b) $2k$ pions created through PV excitation 
with quantum numbers of $A$ intermediate state.
c)~$2k$-pions created by PV excitation with
vacuum quantum numbers intermediate  state.
}
\label{fig1}
\end{figure}

\begin{figure} 
\begin{center}
\includegraphics[scale=1.0]{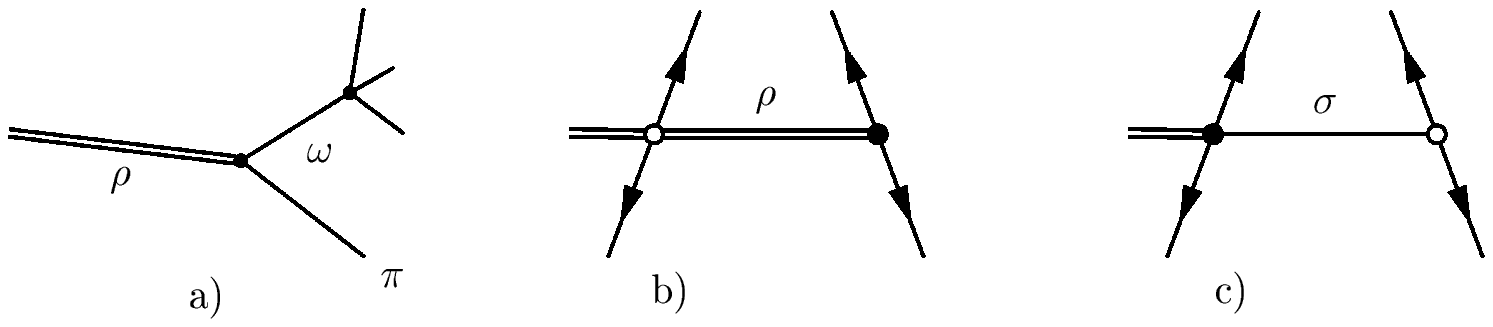}
\end{center}
\caption{  Feynman diagrams for $\rho \to 4 \pi$
decay:
a) without PV excitation,
b,c) two-pion production through PV excitation.
}
\label{fig2}
\end{figure}

\begin{figure} 
\begin{center}
\includegraphics[scale=1.0]{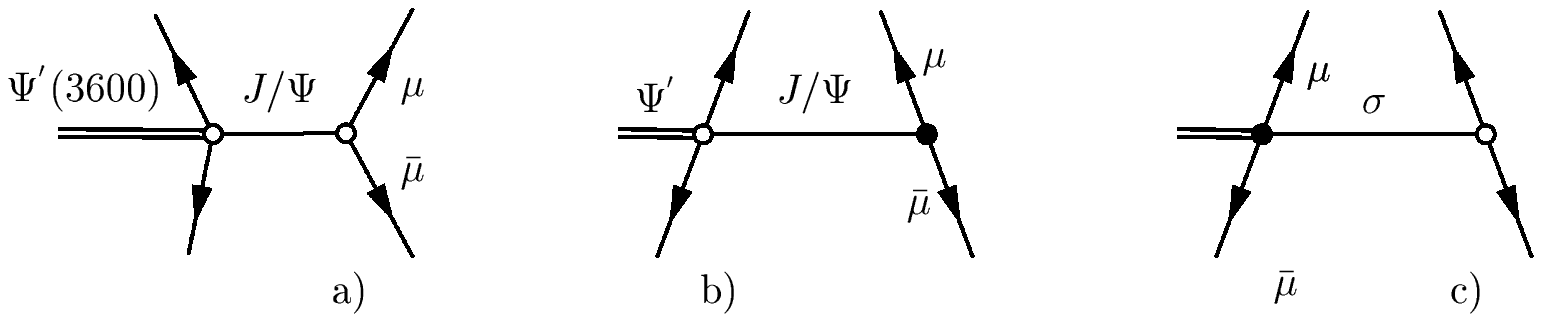}
\end{center}
\caption{ Feynman diagrams for $\Psi ^{'}(3600) \to 2\pi \mu^+ \mu^-$
decay:
a) without PV excitation,
b,c) two-pion created from PV excitation.
}
\label{fig3}
\end{figure}

\end{document}